\documentclass[12pt]{article}
\usepackage{amsfonts,amssymb,amsmath,amsthm}
\usepackage{slashed}
\usepackage{graphicx}
\usepackage{subfig}
\usepackage{mathrsfs}
\usepackage{bbm}
\usepackage{cite}
\usepackage{enumerate}
\usepackage{color}
\usepackage[colorlinks,linkcolor=red,anchorcolor=red,citecolor=green]{hyperref}
\makeatletter

\newcommand{\Rmnum}[1]{\expandafter\@slowromancap\romannumeral #1@}
\makeatother

\begin{document}
\renewcommand{\thefootnote}{\fnsymbol{footnote}}

\vspace{10mm}
\begin{center}
{\Large\bf Analytic phase structures and thermodynamic curvature for the charged AdS black hole in alternative phase space}
\vspace{10mm}

{{\large Zhen-Ming Xu \footnote{E-mail: xuzhenm@nwu.edu.cn}}

\vspace{3mm}
{\em Institute of Modern Physics, Northwest University, Xi'an 710127, China}

\vspace{3mm}
{\em School of Physics, Northwest University, Xi'an 710127, China}

\vspace{3mm}
{\em Shaanxi Key Laboratory for Theoretical Physics Frontiers, Xi'an 710127, China}

\vspace{3mm}
{\em Peng Huanwu Center for Fundamental Theory, Xi’an 710127, China}
}
\end{center}

\vspace{5mm}
\centerline{{\bf{Abstract}}}
\vspace{6mm}
In this paper, we visit the thermodynamic criticality and thermodynamic curvature of the charged AdS black hole in a new phase space. It is shown that when the square of the total charge of the charged black hole is considered as a thermodynamic quantity, the charged AdS black hole also admits an van der Waals-type critical behavior without the help of thermodynamic pressure and thermodynamic volume. Based on this, we study the fine phase structures of the charged AdS black hole with fixed AdS background in the new framework. On the one hand, we give the phase diagram structures of the charged AdS black hole accurately and analytically, which fills up the gap in dealing with the phase transition of the charged AdS black holes by taking the square of the charge as a thermodynamic quantity. On the other hand, we analyse the thermodynamic curvature of the black hole in two coordinate spaces. The thermodynamic curvatures obtained in two different coordinate spaces are equivalent to each other and are also positive. Based on an empirical conclusion under the framework of thermodynamic geometry, we speculate that when the square of charge is treated as an independent thermodynamic quantity, the charged AdS black hole is likely to present a repulsive between its molecules. More importantly, based on the thermodynamic curvature, we obtain an universal exponent at the critical point of phase transition.

\section{Introduction}
It is widely believed that black hole physics is playing a pivotal role in gravity theory. Black hole can hide clues about how to unify general relativity and quantum mechanics. With the pioneering works about the temperature and entropy of black holes\cite{Hawking1975,Bardeen1973,Bekenstein1973,Hawking1983}, thermodynamics theory, which has been rested on general principles spanning a wide range of pure fluids physical systems, is also applied to black hole systems successfully. Now we have known that black holes have rich thermodynamic critical behaviors and microstructures\cite{Wald2001,Padmanabhan2010,Carlip2014,Kubiznak2017,Kastor2009,Dolan2011,Altamirano2014}. The most typical example is the charged AdS black hole, i.e., Reissner-Nordstr\"{o}m AdS (RN-AdS) black hole. Its thermodynamic behavior is very similar to that of the van der Waals fluid in ordinary thermodynamics, and it can show the van der Waals-type phase transition\cite{Kubiznak2012} in the extended phase space. This phenomenon has also been explored in a large number of black hole models with AdS background\cite{Belhaj2015a,Wei2015a,Cai2013,Xu2014,Hendi2016,Odintsov2002,Liu2013,Wang2014,Belhaj2015b,Xu2017,Miao2017,Miao2016,Smailagic2013,Cvetic2011}.

For the RN-AdS black hole, based on the analogy with the related research methods in ordinary thermodynamics, many contents of various thermodynamics or statistical physical properties of the black hole have been expanded. In general, there are two aspects of the researches. Here we first make a brief description of the letter: thermodynamic pressure $P$, thermodynamic volume $V$, temperature $T$, entropy $S$, the total charge $e$ and electrostatic potential $\phi$. On the one hand, at fixed total charge, the thermodynamic criticality and critical exponents in $(P, V)$ plane, i.e., extended phase space, are first shown in literature\cite{Kubiznak2012} and the exactly analytical solutions of the Maxwell construction in $(P, V)$ plane and $(T, S)$ plane are reported in \cite{Spallucci2013}. Furthermore, with the help of the idea of thermodynamics geometry\cite{Ruppeiner1995,Ruppeiner2014,Weinhold1975}, the micro-mechanism of the black hole has been studied phenomenologically\cite{Wei2015,Wei2019,Miao2018,Xu2019a,Ghosh2019}. On the other hand, at fixed AdS background, there are also signs of phase transition in $(e, \phi)$ plane in \cite{Chamblin1999a,Chamblin1999b,Wu2000}. Recently, some works\cite{Dehyadegari2017,Xu2019b} further suggest that when the square of the total charge of the black hole is treated as a thermodynamic quantity, the charged AdS black hole can admit an van der Waals-type critical behavior without the help of thermodynamic pressure $P$ and thermodynamic volume $V$. This new approach toward the critical phenomena of the charged black holes can also work in the Gauss-Bonnet gravity as well as in higher dimensional spacetime\cite{Yazdikarimi2019}.

In the study\cite{Dehyadegari2017}, authors consider $e^2$ as an independent thermodynamic quantity and analyze the critical behavior and microscopic structure of the charged AdS black hole. But so far, there is no exact analytical phase structure in this new framework. In this paper, we try to fill in this gap. Based on regarding $e^2$ as an independent thermodynamic quantity, we study the fine phase structures of the RN-AdS black hole with fixed AdS background. On the one hand, we give the phase diagram of the RN-AdS black hole accurately and analytically. It needs to be explained that in~\cite{Spallucci2013}, authors use the Maxwell construction in $(P, V)$ plane and $(T, S)$ plane to realize the analytical phase structure in the framework of the fixed total charge $e$. While in our present work, we deal with the phase structure of black hole in the case of the fixed AdS background. These two are the analysis of the thermal behavior of black holes in completely different backgrounds. In the framework of the fixed AdS background and the square of charge $e^2$ as an independent thermodynamic quantity, we obtain a new and analytic phase structure behavior of the charged AdS black hole. On the other hand, we analyse the thermodynamic curvature of the black hole in two coordinate spaces and obtain an universal exponent at the critical point of phase transition. In addition, we speculate that when the square of charge is treated as an independent thermodynamic quantity, our obtained positive thermodynamic curvature may be considered as a probe of the interaction of repulsive between the RN-AdS black hole molecules, based on an empirical conclusion under the framework of thermodynamic geometry theory.

The paper is organized as follows. In section \ref{sec2}, we briefly review some basic thermodynamic properties of the RN-AdS black hole and give a sufficient reason to introduce a new pair of conjugate quantities. In section \ref{sec3}, we study the fine phase structures of the RN-AdS black hole with fixed AdS background. In section \ref{sec4}, we have calculated the thermodynamic curvature in two different coordinate spaces. Finally, we devote to drawing our conclusion in section \ref{sec5}. Throughout this paper, we adopt the units $\hbar=c=k_{_{B}}=G=1$.

\section{Basic thermodynamic quantities}\label{sec2}
The metric of the RN-AdS black hole is\cite{Kubiznak2017,Kubiznak2012}
\begin{equation}
d s^2=-f(r)dt^2+\frac{d r^2}{f(r)}+r^2(d\theta^2+\sin^2 \theta d\varphi^2),
\end{equation}
and here the function $f(r)$ is
\begin{equation}\label{laspon}
f(r)=1-\frac{2M}{r}+\frac{r^2}{l^2}+\frac{e^2}{r^2},
\end{equation}
where the parameter $M$ represents the ADM mass of the black hole, $e$ stands for the total charge and $l$ is the curvature radius of the AdS spacetime. Some basic thermodynamic properties of the RN-AdS black hole take the following forms in terms of the event horizon radius $r_h$, which is regarded as the largest root of equation $f(r)=0$\cite{Kubiznak2017,Kubiznak2012},
\begin{eqnarray}
\text{Enthalpy}&:&M=\frac{r_h}{2}+\frac{r_h^3}{2l^2}+\frac{e^2}{2r_h},\label{enthalpy}\\
\text{Temperature}&:&T=\frac{1}{4\pi r_h}+\frac{3r_h}{4\pi l^2}-\frac{e^2}{4\pi r_h^3},\label{temperature}\\
\text{Entropy}&:&S=\pi r_h^2,\label{entropy}\\
\text{Pressure}&:&P=\frac{3}{8 \pi l^2 },\label{pressure}\\
\text{Thermo-volume}&:&V=\frac{4}{3} \pi r_h^3.\label{volume}
\end{eqnarray}

A new pair of conjugations are thermal-charge $Q$ and thermal-potential $\Psi$ in \cite{Xu2019b},
\begin{equation}\label{thermalcharge}
Q=e^2, \qquad \Psi=\left(\frac{\partial M}{\partial Q}\right)_{S,P}=\frac{1}{2r_h}.
\end{equation}
The main reasons for the introduction of the above new conjugate quantity are as follows.
\begin{itemize}
  \item Based on the theory of the extended phase space, one can relate the square of the curvature radius of the AdS spacetime $l^2$ as the thermodynamic pressure~(\ref{pressure}). By observing the form of solution of the RN-AdS black hole, we can see that the charge always appears in the form of $e^2$. First, in order to maintain formal consistency, we can regarded $e^2$ as an independent thermodynamic quantity. Second, once the $e^2$ as an independent thermodynamics quantity, its conjugate quantity $\Psi$ is the simplest, which means that it is only a function of the horizon radius $r_h$. For the electric potential $\phi=q/r_h$, it is a function of $q$ and $r_h$.
  \item Although the new physical quantity $\Psi$ is different from the electric potential $\phi$, their contributions to the internal energy of the system are the same, i.e., $2Q\Psi=q\phi$. In addition, according to the idea of the black hole micromolecule in \cite{Wei2015}, the $\Psi$ is the number density of black hole micromolecules in the natural units, and also it is the inverse of the specific volume defined in \cite{Kubiznak2012}.
  \item Completely from the perspective of thermodynamics, Refs.~\cite{Aman2003,Aman2006a,Aman2006b,Mirza2007} have shown that the charged black hole is Ruppeiner flat, which means that the black hole is a non-interacting statistical system. While this suggestion is inconsistent with our intuitive understanding of the charged black hole. There should be electromagnetic interaction between its molecules. The research\cite{Xu2019b} shows that the charged black hole is Ruppeiner curve (i.e., an interacting statistical system) by treating $e^2$ as an independent thermodynamic quantity.
  \item  When $e^2$ is an independent thermodynamic quantity, the black hole also exhibits the van der Waals-type phase transition behavior in the new phase space, without the help of the thermodynamic pressure $P$ and thermodynamic volume $V$. Meanwhile, we find that the phase transition can be solved accurately. This is a very unexpected and interesting result because the van der Waals-type phase transition of black holes basically appeared in the extended phase space with the thermodynamic pressure $P$ and thermodynamic volume $V$. In the framework of the square of charge $e^2$ as an independent thermodynamic quantity, we find that the charged AdS black hole exhibit an \emph{abnormal} phase structure behavior. In general, for the van der Waals-type thermodynamic system, the critical region is located at $T<T_c$, and when $T>T_c$ it is called supercritical region. For the scheme we are dealing with at present, it is just the opposite, i.e., $T>T_c$ for the critical region while $T<T_c$ for supercritical region. This may just show some unique features of black hole thermodynamic system different from ordinary thermodynamic system.
  \item Although the black hole thermodynamics has made a good development, whether the black hole system is a real thermodynamic system or just a thermodynamic system in form is still in the exploratory stage so far. Therefore, when we deal with the thermodynamics of black holes, there is no strict structure for the expression of the first law of thermodynamics, and the role of the parameters of black holes in the first law seems vague. Various attempts have been made to solve these problems, like that the cosmological constant~\cite{Kastor2009}, the Born–Infeld parameter~\cite{Gunasekaran2012,Breton2005}, the Gauss–Bonnet coupling constant~\cite{Cai2013} or the reciprocal of the Gauss–Bonnet coupling constant~\cite{Xu2014}, the Horndeski coupling constant~\cite{Miao2016b} and the noncommutative parameter~\cite{Miao2017b} can be dealt with as a kind of thermodynamic pressure. The same is true for the charge. If the charge itself is regarded as an independent thermodynamic quantity, then its conjugate quantity is the electric potential. If the square of the charge is treated as an independent thermodynamic quantity, as explained in the first point above, its conjugate quantity is number density of black hole micromolecules. Although the two forms are different, their contribution to the total mass of the black hole is the same.
\end{itemize}

Hence the first law of thermodynamics and Smarr relation can be written in terms of the thermodynamic quantities mentioned above as follows:
\begin{equation}\label{fsrealtion}
d M=T d S+V dP+\Psi dQ, \qquad M=2(TS-PV+\Psi Q).
\end{equation}

\section{The fine phase structures}\label{sec3}

Next we consider the criticality of the RN-AdS black hole at fixed pressure $P$, i.e., fixed AdS background. In $(T,S)$ plane, the equation of state of the RN-AdS black hole is
\begin{equation}\label{state1}
T=\frac{1}{4\sqrt{\pi S}}+\frac{2P\sqrt{S}}{\sqrt{\pi}}-\frac{\sqrt{\pi}Q}{4S\sqrt{S}},
\end{equation}
and in $(Q,\Psi)$ plane, the equation of state of the RN-AdS black hole is
\begin{equation}\label{state2}
Q=\frac{1}{4\Psi^2}+\frac{\pi P}{2\Psi^4}-\frac{\pi T}{2\Psi^3}.
\end{equation}

For the analysis of the criticality of the black hole, the most important thing is to calculate the exact phase diagram equation by means of the Maxwell construction. We now make detailed calculation in two planes $(T,S)$ and $(Q,\Psi)$ respectively. Correspondingly the critical values satisfy equations $\partial T/\partial S=0=\partial^2 T/\partial S^2$ at critical thermal-charge $Q_c$ or $\partial Q/\partial \Psi=0=\partial^2 Q/\partial \Psi^2$ at critical temperature $T_c$ . Hence we have
\begin{equation}\label{cvalues}
Q_c=\frac{1}{96\pi P},\quad \Psi_c=2\sqrt{\pi P},\quad T_c=\frac43\sqrt{\frac{P}{\pi}}, \quad S_c=\frac{1}{16P}.
\end{equation}
For convenience, we first introduce the following dimensionless symbols
\begin{equation}\label{symbols}
t=\frac{T}{T_c}, \quad q=\frac{Q}{Q_c}, \quad z=\frac{\Psi}{\Psi_c}, \quad s=\frac{S}{S_c},
\end{equation}
and the equation of state~(\ref{state1}) and~(\ref{state2}) can be rewritten as
\begin{equation}\label{rstate2}
t=\frac{3}{4\sqrt{s}}+\frac{3\sqrt{s}}{8}-\frac{q}{8s\sqrt{s}},
\end{equation}
and
\begin{equation}\label{rstate2}
q=\frac{6}{z^2}+\frac{3}{z^4}-\frac{8t}{z^3}.
\end{equation}

In $(T,S)$ (or $(t,s)$) plane, according to Maxwell construction, we have
\begin{eqnarray}
  && t|_{s=s_1}=t|_{s=s_2}=t^{*}, \nonumber \\
  && t^{*}\cdot(s_2-s_1)=\int_{s_1}^{s_2}tds,
\end{eqnarray}
where $t^{*}$ stands for an isotherm, $s_1$ and $s_2$ denote the entropy of the small and large black hole phases respectively. By solving the above equations, we finally get
\begin{eqnarray}\label{phase1}
  && t^{*}=\sqrt{\frac{3-\sqrt{q}}{2}}, \nonumber \\
  && s_{1,2}=\frac14\left(\sqrt{6-2\sqrt{q}}\pm \sqrt{6-6\sqrt{q}}\right)^2.
\end{eqnarray}

In $(Q,\Psi)$ (or $(q,z)$) plane, the Maxwell construction impies
\begin{eqnarray}
  && q|_{z=z_1}=q|_{z=z_2}=q^{*}, \nonumber \\
  && q^{*}\cdot(z_2-z_1)=\int_{z_1}^{z_2}qdz,
\end{eqnarray}
where $q^{*}$ stands for the constant $q$ curve, $z_1$ and $z_2$ denote the thermal-potential of the large and small black hole phases respectively. So we finally obtain
\begin{eqnarray}\label{phase2}
  && q^{*}=(3-2t^2)^2, \nonumber \\
  && z_{1,2}=\frac{1}{t\pm \sqrt{3t^2-3}}.
\end{eqnarray}

\begin{figure}
\begin{center}
\includegraphics[width=110mm]{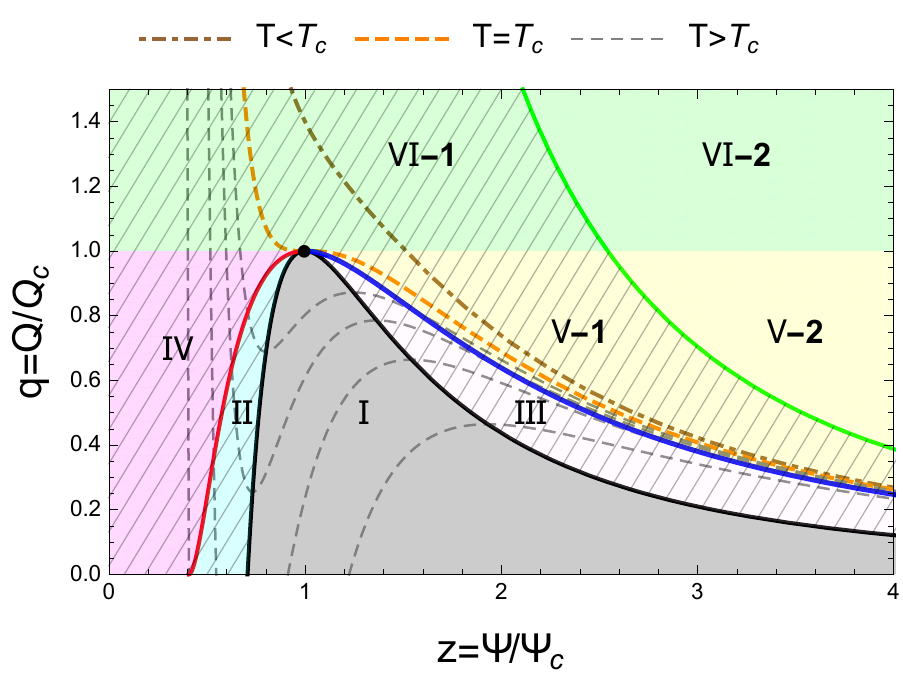}
\end{center}
\caption{(color online) The fine phase structures of the RN-AdS black hole in $(Q,\Psi)$ plane.}
\label{fig1}
\end{figure}

In fact, the phase diagram equations in these two planes are equivalent, i.e., Eqs.~(\ref{phase1}) and~(\ref{phase2}) are equivalent. Therefore, we only need to analyze one of them. We take the case in $(Q,\Psi)$ plane as an example, the fine phase structures of the RN-AdS black hole are shown in Figure \ref{fig1}.
Next, we will detail the phase structure one by one.
\begin{description}
  \item[Critical and supercritical regions] In region \Rmnum{1}, \Rmnum{2}, \Rmnum{3}, \Rmnum{4} and \Rmnum{5}, we can clearly observe that there is oscillation in the curve of the equation of state, and there is a local maximum and a local minimum in the curve, see gray dashed line. At critical point (see black point), the two local extremes are merged into one (see orange dashed line). Hence we obtain that the critical region contains the five regions just mentioned, which satisfies $0<q\leq1$ and $t\geq 1$. In the region \Rmnum{6}, we have $q>1$ and $0<t<1$. The oscillation in the equation of state disappears, see brown dot-dashed line. This is the supercritical region in which there is no phase separation and no physical distinction between small and large black hole phases. We notice that the result here are the opposite of that of fixed total charge in $(P, V)$ plane or $(T, S)$ plane, where one can have $0<t<1$ for the critical region and $t>1$ for the supercritical region\cite{Spallucci2013}.
  \item[Single-phase and coexistence regions] At $0<q\leq1$ and $t\geq 1$, we know there is van der Waals-type phase transition. We must utilize the usual Maxwell construction to remove the oscillating part in the curve of the equation of state. Thus the whole black hole system can transition from the large black hole phase to the small black hole phase through the constant $q$ process. Therefore the two-phase coexistence regions are \Rmnum{1}, \Rmnum{2} and \Rmnum{3}. The single-phase of large black hole region is \Rmnum{4} and the single-phase of small black hole region is \Rmnum{5}. The boundary of the single-phase of large black hole region and the coexistence region, i.e., red solid line, satisfies parametric curve equation
      \begin{equation}\label{lcb}
      q=(3-2t)^2, \qquad z=\frac{1}{t+\sqrt{3t^2-3}},
      \end{equation}
      and the boundary of the single-phase of small black hole region and the coexistence region, i.e., blue solid line, satisfies parametric curve equation
      \begin{equation}\label{scb}
      q=(3-2t)^2, \qquad z=\frac{1}{t-\sqrt{3t^2-3}}.
      \end{equation}

      In the coexistence region, the region \Rmnum{2} corresponds to superheated large black hole while the region \Rmnum{3} is supercooled small black hole. Both of them are in metastable state. In these two regions, the single phase transits to two-phase coexistence equilibrium state through nucleation growth. Small black hole formed in superheated large black hole and large black hole formed in supercooled small black hole due to fluctuation. While for region \Rmnum{1}, it is completely unstable and the transition mode to realize two-phase coexistence equilibrium state is called spinodal decomposition. The boundary (black solid line) of spinodal decomposition region satisfies Eq.~(\ref{rstate2}) and $\partial q/\partial z=0$, i.e.,
      \begin{equation}\label{spb}
      q=\frac{2}{z^2}-\frac{1}{z^4}.
      \end{equation}
  \item[Stable and unstable regions] We know that the thermodynamic stability of the black hole is determined by the heat capacity. For the fixed AdS background situation we are discussing at present, we have the heat capacity at constant $q$ and the heat capacity at constant $z$, that is $c_{q,z}:=t(\partial s/\partial t)_{q,z}$,
      \begin{equation}\label{capacity}
      c_q=\frac{2(qz^2-3z^{-2}-6)}{3(2z^2-qz^4-1)}, \qquad c_z=0.
      \end{equation}
      It is very similar to heat capacity at constant pressure $c_p$ and heat capacity at constant volume $c_v$ of black holes (for RN-AdS black hole we have $c_p\neq 0, ~c_v=0$.). Hence we obtain that the black hole is locally stable for $c_q>0$, but unstable for $c_q<0$. For the heat capacity at constant $q$, the numerator corresponds to the green solid line, and the denominator corresponds to the black solid line. We can observe that unstable regions are \Rmnum{1}, \Rmnum{5}-2 and \Rmnum{6}-2. The rest (\Rmnum{2}, \Rmnum{3}, \Rmnum{4}, \Rmnum{5}-1 and \Rmnum{6}-1) are stable regions, i.e., the oblique line area. We notice that regions \Rmnum{2} and \Rmnum{3} fall within this range. In fact, the two regions are in metastable state. Therefore, there are still some subtleties in the analysis of the stability of the black hole system from the perspective of heat capacity alone. We have to rely on more fine phase structures to get a reasonable answer.
\end{description}

\section{Thermodynamic curvature}\label{sec4}
Now let us consider the Ruppeiner thermodynamic geometry which is based on the Hessian matrix about the black hole entropy, and its line element is
\begin{equation}\label{rmetric}
\Delta l^2=-\frac{\partial^2 S}{\partial X^{\mu}\partial X^{\nu}} \Delta X^{\mu} \Delta X^{\nu},
\end{equation}
where $X^{\mu}$ represents some independent thermodynamic quantities. For the charged AdS black hole, in general, it is difficult to express entropy as an analytic function of other thermodynamic quantities. Hence we need to find other equivalent expression of the metric~(\ref{rmetric}). For our current research situation, namely with fixed AdS background and in the new coordinate space $(Q,\Psi)$, we choose the following two ways to analyze the thermodynamic curvature behavior\footnote{In principle, there should be four ways, i.e. in coordinate spaces $\{S,Q\}$, $\{T,\Psi\}$, $\{S,\Psi\}$ and $\{T,Q\}$. Because entropy $S$ and thermal-potential $\Psi$ are not independent of each other, the coordinate space $\{S,\Psi\}$ is invalid. In addition, in coordinate space $\{T,Q\}$, we need to write the entropy $S$ as a function of temperature $T$ and thermal-charge $Q$. Its analytical form is very complex. For the sake of simplicity, we will not consider this case.}.

In the coordinate space $\{S,Q\}$, we have
\begin{equation}\label{lines}
\Delta l^2=\frac{1}{T}\left(\frac{\partial T}{\partial S}\right)_Q \Delta S^2+\frac{2}{T}\left(\frac{\partial T}{\partial Q}\right)_S \Delta S \Delta Q,
\end{equation}
and in the coordinate space $\{T,\Psi\}$, we obtain
\begin{equation}\label{linet}
\Delta l^2=\frac{2}{T}\left(\frac{\partial S}{\partial \Psi}\right)_T \Delta T \Delta \Psi+\frac{1}{T}\left(\frac{\partial Q}{\partial \Psi}\right)_T \Delta \Psi^2.
\end{equation}
With the help of Eqs.~(\ref{entropy}),~(\ref{thermalcharge}),~(\ref{state1}) and~(\ref{state2}), we finally get the thermodynamic curvature in the two coordinate spaces
\begin{eqnarray}\label{tr}
R_{SQ}=\frac{1+16PS}{8PS^2+S-\pi Q} \qquad \text{and} \qquad R_{T\Psi}=\frac{2\Psi(4\pi P+\Psi^2)}{\pi^2 T}.
\end{eqnarray}

There is an empirical observation that negative (positive) thermodynamic curvature is associated with attractive (repulsive) microscopic interactions for a thermodynamic system. Although, because of the absence of a hitherto underlying theory of quantum gravity, there has been a controversial discussion on whether the characterizations of thermodynamic scalar curvature can be directly generalized to the black hole thermodynamics or not\cite{Dolan2015}.  But the well-established black hole thermodynamics makes the Ruppeiner thermodynamic geometry be plausible to phenomenologically or qualitatively provide the information about interactions of black holes. Hence it is also believed that there is such an observation for black holes. Back to our study, we can clearly see that $R_{SQ}=R_{T\Psi}>0$. The positive thermodynamic curvature may be related to the information of repulsive interaction between black hole molecules for the charged AdS black hole with fixed AdS background and treating the square of the charge as a thermodynamic quantity.

When the phase transition occurs, with the help of Eqs.~(\ref{cvalues}), ~(\ref{symbols}), ~(\ref{phase2}) and~(\ref{tr}), we can obtain the thermodynamic curvature of the large and small black hole phases respectively
\begin{eqnarray}
 \widetilde{R}_{1,2}=\frac{R_{1,2}}{R_c}=\frac{z_{1,2}(z^2_{1,2}+1)}{2t},
\end{eqnarray}
where $R_c=24P$ is the critical thermodynamic curvature. When crossing the two-phase coexistence curve, the difference between the thermodynamic curvature of the large and small black hole phases is
\begin{eqnarray}
\Delta\widetilde{R}=\widetilde{R}_2-\widetilde{R}_1=4 \sqrt{6}(t-1)^{1/2}+49 \sqrt{6} (t-1)^{3/2}+\mathcal{O}\left[(t-1)^{5/2}\right].
\end{eqnarray}
We now define a new critical exponent and its corresponding critical amplitude, which is related to the difference between the thermodynamic curvature of the small and large black hole on crossing the coexistence curve
\begin{eqnarray}
\Delta\widetilde{R}=h (t-1)^\zeta.
\end{eqnarray}
Hence we have
\begin{eqnarray}
\zeta=\frac12, \qquad h=4 \sqrt{6}.
\end{eqnarray}
The above critical exponent $\zeta=1/2$ describes the behavior of thermodynamic curvature when the temperature approaching phase transition point and it does not depend on the details of physical systems, i.e., it is universal. While the above critical amplitude $h=4 \sqrt{6}$ depends on the details of physical systems, i.e., it is a characteristic quantity. Similar phenomena also occur in five-dimensional neutral Gauss-Bonnet AdS black hole~\cite{Miao2018b}, in which the critical exponent and its corresponding critical amplitude read as $\zeta=1/2$ and $h=6 \sqrt{6}$.

\section{Summary}\label{sec5}
With the help of the Maxwell construction, at fixed AdS background, we exactly give the phase diagram equations of the charged AdS black hole in $(T,S)$ and $(Q,\Psi)$ plane respectively, see Eqs.~(\ref{phase1}) and~(\ref{phase2}), by treating the square of the total charge of the black hole as a thermodynamic quantity. By means of these exact analytical phase diagram equations, we describe the fine phase structures of the RN-AdS black hole in detail in Figure \ref{fig1}. The whole phase diagram is divided into three parts: critical and supercritical regions, single-phase and two-phase coexistence regions, stable and unstable regions. The boundary of the single-phase region and the coexistence region is shown in Eqs.~(\ref{lcb}) and~(\ref{scb}). For the coexistence region, we further refine it into nucleation growth and spinodal decomposition, both of which are the transition mechanism from single-phase state to two-phase coexistence equilibrium state. Meanwhile we exactly give the boundary equation of spinodal decomposition, see Eq.~(\ref{spb}). In addition, according to the thermodynamic stability analysis based on heat capacity, we point out that this may not be comprehensive and reasonable. In fact we will mistake metastable region for stable region. Therefore, when analyzing the thermodynamic stability of black hole, we still need to use more fine phase structures to get a reasonable answer.

Based on the Ruppeiner thermodynamic geometry, we calculate the thermodynamic curvature in two different coordinate spaces, $\{S,Q\}$ and $\{T,\Psi\}$ in detail. The results show that the thermodynamic curvatures~(\ref{tr}) obtained in the two coordinate spaces are equivalent. The obtained thermodynamic curvature is always positive which may be related to the information of repulsive interaction between black hole molecules for the charged AdS black hole with fixed AdS background and the square of the total charge as a thermodynamic quantity. Meanwhile we obtain a universal exponent $\zeta=1/2$ at the critical point of phase transition.

In short, in this paper the accurate and analytical phase diagram structures of the RN-AdS black hole in the new phase space enrich the researches about the RN-AdS black hole, making the content of the criticality and phase transition of the charged black hole more substantial and complete. Meanwhile our results fill up the gap in dealing with the phase transition of charged AdS black holes by taking the square of the charge as a thermodynamic quantity. This approach should also be applied to other black hole models, like the charged Gauss-Bonnet gravity as well as in higher dimensional spacetime.

\section*{Acknowledgments}
The financial supports from the National
Natural Science Foundation of China (Grant Nos. 11947208 and 11947301), China Postdoctoral Science Foundation (Grant No. 2020M673460) are gratefully acknowledged. This research is also supported by The Double First-class University Construction Project of Northwest University. The authors would like to thank the anonymous referee for the helpful comments that improve this work greatly.

\end{document}